\pdfoutput=1 

\documentclass[twocolumn,10pt]{article}
\usepackage{amsfonts}
\usepackage{amsmath}
\usepackage{amssymb}
\usepackage{amsthm}
\usepackage[sort]{natbib}
\bibpunct{[}{]}{,}{a}{}{,}  % Use this natbib punctuation
\usepackage[dvips]{graphicx}
\usepackage[usenames,dvipsnames]{color}
\usepackage{colortbl}
\usepackage{wasysym} 			% contains \smiley and \frownie
\usepackage{subfigure}
\usepackage{url}
\usepackage[bookmarks, colorlinks=false]{hyperref}
\usepackage[labelfont=bf,font={sf}]{caption} % makes caption label bold-sf and text sf
\usepackage{verbatim}
\usepackage{ulem}			% better underlining

\def\figs{Figures}                              % path to image files
\newcommand{\ssection}[1]{%
     \section[#1]{\sf \large \textbf{#1}}}
\newcommand{\ssubsection}[1]{%
     \subsection[#1]{\sf \textbf{#1}}}

\newtheorem{thm}{Theorem}

\newtheorem{cor}{Corollary}

\newtheorem{prf}{Proof}

\begin{document} 
\sffamily %sets global text to sans serif font (but doesn't affect section titles)

%---------------------------------------------------------------------------
%%                          TITLE
%% Text is 10pt, then ...
%% Title should be 14pt == Large
%% Names should be 12pt == large
%---------------------------------------------------------------------------
\title{\vspace{-0.25in} \huge \bfseries \textsf{Getting in the Zone for Successful Scalability}\footnotemark[2]
}
\author{\Large Jim Holtman$^a$ and Neil J. Gunther$^b$  \\
{\large Cincinnati, Ohio, USA$^a$} \\
{\large Performance Dynamcis Company, Castro Valley, California, USA$^b$}  \\
{\normalsize \tt \{jholtman@gmail.com, njgunther@perfdynamics.com\} } \\
%\author{\large Paper 7050 \\
%{\normalsize Draft of \today. Submitted to CMG 2007}
}
\date{}

\twocolumn[
    \begin{@twocolumnfalse}
      \maketitle % this MUST go no earlier than here!
      \begin{center}
      \parbox{5.5in}{\large %to match the layout instructions
		\em  The universal scalability law (USL) is an analytic model
		used to quantify application scaling. It is universal because it
		subsumes Amdahl's law  and Gustafson linearized scaling as
		special cases. Using simulation, we show: (i) that the USL is
		equivalent to synchronous queueing in a load-dependent machine
		repairman model and (ii) how USL, Amdahl's law and Gustafson
		scaling can be regarded as boundaries defining three scalability
		zones. Typical throughput measurements lie across all three zones.
		Simulation scenarios provide deeper insight into queueing
		effects and thus provide a clearer indication of which
		application features should be tuned to get into the optimal
		performance zone.
        }
        \end{center}
      \vspace{20pt}
    \end{@twocolumnfalse}
  ]
  {
    \renewcommand{\thefootnote}%
    {\fnsymbol{footnote}}
    \footnotetext[2]{\sf~Copyright \copyright ~2008 Gunther, Holtman. All
    Rights Reserved. This document may not be reproduced, in whole or in
    part, by any means, without the express permission of the authors.
    Permission has been granted to CMG, Inc. to publish in the Proceedings
    and the associated CD. Revised~\today}
  }
\thispagestyle{empty} % enforce no number on 1st page

%% http://www1.umn.edu/urelate/style/language-usage.html
%% Orient, orientate. 
%% Orientate has crept into the language, probably as a
%% back-formation from orientation, but it is a superfluous word. Save a
%% syllable and use orient.

%% Starts with first section 
%% This is input into main document source.

%% http://www1.umn.edu/urelate/style/language-usage.html
%% Orient, orientate. 
%% Orientate has crept into the language, probably as a
%% back-formation from orientation, but it is a superfluous word. Save a
%% syllable and use orient.

\ssection{INTRODUCTION} \label{sec:intro}
The 2008 JavaOne conference included many presentations on techniques for 
achieving better scalability, e.g., caching, collocation,
 ``parallelization,'' and pooling. But these are only {\em
qualitative} descriptions. How can the impact of applying such
techniques be \uline{quantified}? Clearly, this is the domain of 
performance modeling. Performance models are essential, not only for
prediction, but also for {\em interpreting} scalability measurements.
However, most performance modeling tools, e.g., PDQ~\citep{ppdq} and
SIMUL8~\citep{holtsim}, use a queueing paradigm which requires measured
service times as {\em inputs} to parameterize each  queueing
facility being modeled. More often than not, such detailed measurements
are not available, thereby thwarting this approach.

\begin{figure}[!htb]
\centering
\includegraphics[scale = 0.65]{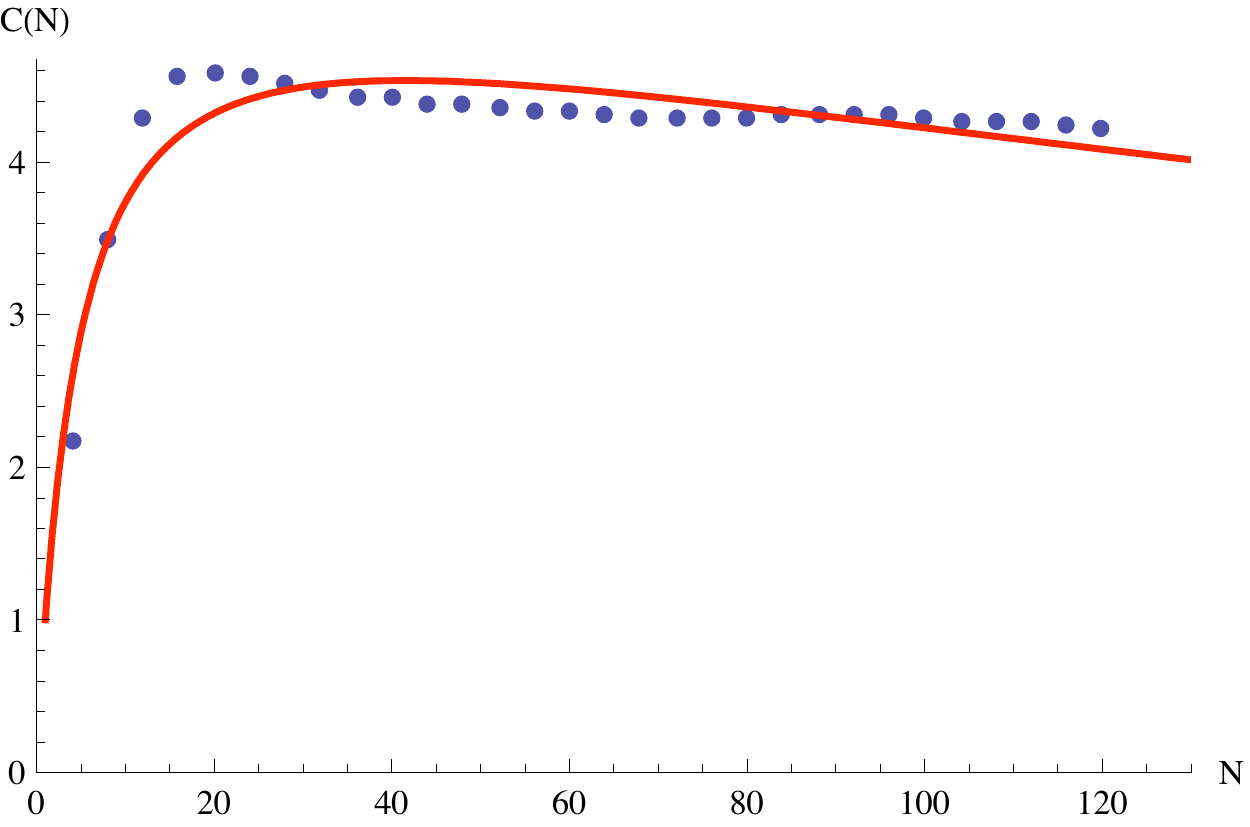} 
\caption{USL parametric model (red) fitted to Websphere relative capacity
measurements $C(N)$ as a function of user load $N$ with coefficients
\mbox{$\alpha=0.18169$} and \mbox{$\beta=0.00047$}. Retrograde
performance is clear. Amdahl and Gustafson parametric models cannot
accommodate this effect}.
\label{fig:USLwebsphere}
\end{figure}

A more practical approach is based on statistical
regression~\citep{holtR} of measured throughput data using {\em
parametric models}; the advantage being that service-time measurements
are not required. One parametric model that has been used successfully
for the past two decades is based on the {\em  universal scalability
law} or USL~\cite{cmg93,gcap,arxiv08}. A distinguishing feature of the
USL is its ability to analytically model the {\em retrograde} throughput
(Fig.~\ref{fig:USLwebsphere}) commonly seen in custom benchmarks and
load-test measurements. If such retrograde behavior is not present, the
USL reduces to either Amdahl's law (See Sect.~\ref{sec:paramamd}) or
Gustafson's linearized form (See Sect.~\ref{sec:paramgus}).

As useful as all this has been, the question has remained: Do parametric
models like USL represent something more fundamental? We answer that
question in the affirmative by showing that the USL corresponds to a
certain bounding curve on the throughput of a {\em machine repairman}
(MRM) queueing model~\citep{gross}. We use event-based simulation as an
exploratory tool to investigate the precise conditions under which such
a bound can exist. Amdahl's law and Gustafson's linearization are
contained as special cases of the MRM model. Based on this new insight,
we introduce the concept of {\em scalability zones}. Rather than relying
on any particular bounding curve to express scalability, we show how
each of these bounding curves defines a set of zones. In practice,
typical throughput measurements lie across all these zones and this
provides a more helpful interpretation for determining potential 
performance improvements.

Our paper is organized as follows. In Sect.~\ref{sec:parametrics} we
review each of the parametric scalability models discussed subsequently.
In Sect.~\ref{sec:qmodels} we review the repairman model and the generalizations
necessary to make contact with the parametric models. These extensions
include: (i) a prepping repairman in Sect.~\ref{sec:prepman} and (ii)
synchronous queueing in Sect.~\ref{sec:synq}. Sect.~\ref{sec:sims}
describes the simulation models and presents the results that support
the identification of the USL with synchronous queueing at a prepping
repairman, together with several special cases. Finally, in
Sect.~\ref{sec:zones}, we discuss a new approach to interpreting
scalability data in terms of {\em zones} and transitions between them.
These zones are well-defined in terms of queueing effects and thus, can
provide vital insights into how best to improve scalability. An example
of applying the zone concept to actual scalability measurements is
presented.

\ssection{PARAMETRIC MODELS} \label{sec:parametrics}
We begin our analysis by defining the {\em relative capacity}:
\begin{equation}
C(N) = X(N)/X(1) \label{eqn:relcap}
\end{equation}
where $X(N)$ represents the throughput generated by either, $N$
processors in the case of hardware scalability~\citep[See][Chap.
4]{gcap} or $N$ virtual users in the case of software
scalability~\citep[See][Chap. 6]{gcap}. The ratio in (\ref{eqn:relcap})
has two possible interpretations:
\begin{description}\addtolength{\itemsep}{-0.5\baselineskip}
\item[Data representation:] 
$X(N)$ represents the actual throughput
measurements, e.g., transactions per second. The relative capacity
$C(N)$ is simply the {\em normalization} of those data. See Fig.~\ref{fig:USLwebsphere}.
\item[Analytic representation:] 
$X(N)$ is represented by a
function, e.g., a linear regression model such as: $X(N) = mN+c$ where
$m$ is the {\em slope} and $c$ is the {\em intercept}. 
See Sect.~\ref{sec:simgus}
\end{description}
With regard to each of the parametric models discussed in this section,
we shall demonstrate that $C(N)$ is best represented,
not by a simple linear function, but by a ratio of functions:
  \begin{equation}
C(N) = P(N)/Q(N) \label{eqn:ratfun}
\end{equation}
where $P(N)$ and $Q(N)$ are polynomials in $N$.
Such functions are called {\em rational functions}
(See \url{en.wikipedia.org/wiki/Rational_function}).

\begin{figure}[!htb]
\centering
\includegraphics[scale = 0.65]{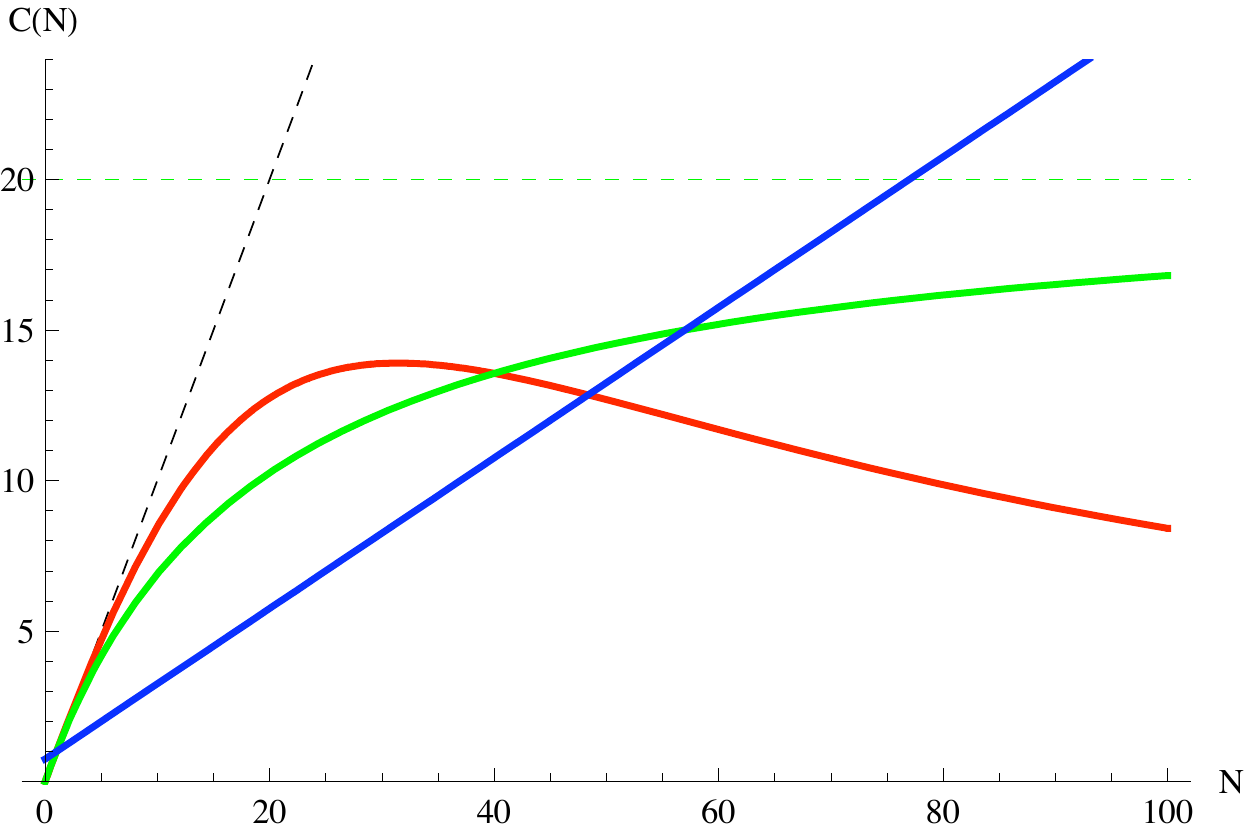} 
\caption{Parametric models: USL (red), Amdahl (green), Gustafson (blue),
with  parameter values exaggerated to distinguish their typical
characteristics relative to ideal linearity (dashed black). The
horizontal dashed green line is the Amdahl asymptote at $\alpha^{-1}$.
Compare with the application of the USL model in Fig.~\ref{fig:USLwebsphere}}.
\label{fig:funcmodels}
\end{figure}  

We pause to reflect on the significance of (\ref{eqn:ratfun}). Computer
system scalability can be modeled using many possible functions which are
{\em not} rational functions, e.g., geometric scaling~\citep{cmg96}. As
we shall show in Sects.~\ref{sec:qmodels} and~\ref{sec:sims}, rational
functions are physically correct because they possess a deep connection
with queueing theory. As far as we are aware, this fundamental
relationship between parametric models and queueing models
has not been elucidated before. Conversely, geometric scaling can be
excluded on the grounds that it is unphysical from
the standpoint of queueing theory~\citep{gcap,arxiv02}.

\ssubsection{Universal Scalability Law} \label{sec:paramusl}
The most general parametric model of scalability 
is the two-parameter universal scalability law (USL):
\begin{equation}
C(N,\alpha,\beta) = \dfrac{N}{1 + \alpha (N-1) + \beta N (N-1)} \label{eqn:usl}
\end{equation}
which is a rational function with $P(N)=N$ and \mbox{$Q(N)=aN^2 + bN +
c$}; a quadratic polynomial with coefficients $a,b,c \geq 0$. These
coefficients have been regrouped into three terms involving only two
parameters $\alpha, \beta  \geq 0$, in the denominator of (\ref{eqn:usl}). These 
three terms can be interpreted as the ``Three C's'':

\begin{enumerate}\addtolength{\itemsep}{-0.5\baselineskip}
\item \fbox{C}oncurrency-limited scalability when  
\mbox{$\alpha,\beta = 0$} such that $C(N) \sim N$, i.e., linear scaling.

\item \fbox{C}ontention-limited scalability due to serialization or queueing,  
i.e., when \mbox{$\alpha > 0, \beta = 0$}.

\item \fbox{C}oherency-limited scalability due to the delay incurred by making local copies 
of data or instructions consistent across multiple caches or nodes, 
i.e., when \mbox{$ \alpha, \beta > 0$}.
\end{enumerate}

Table~\ref{tab:appclass} summarizes how these parameter values can be used to  
classify the scalability of different types of platforms and applications.

\begin{table} 
\caption{Application classes for the USL model} \label{tab:appclass}
\centering
\begin{tabular}{c|l} 
\hline
			& {\bf Ideal concurrency} ($\alpha,\beta = 0$)\\ \cline{2-2}
			& Single-threaded tasks\\	 
\bf A		& Parallel text search\\ 
			& Read-only queries\\	 
\hline
			& {\bf Contention-limited} ($\alpha > 0, \beta = 0$)\\ \cline{2-2}
			& Tasks requiring locking or sequencing\\
\bf B	 	& Message-passing protocols\\
			& Polling protocols (e.g., hypervisors)\\
\hline
			& {\bf Coherency-limited} ($\alpha = 0, \beta > 0$)\\ \cline{2-2}
			& SMP cache pinging\\
\bf C		& Incoherent application state between\\
			& cluster nodes\\
\hline
			& {\bf Worst case} ($\alpha, \beta > 0$)\\ \cline{2-2}
			& Tasks acting on shared-writable data \\
\bf D		& Online reservation systems \\
			& Updating database records  \\
\hline
\end{tabular}
\end{table}

\ssubsection{Amdahl's  Law} \label{sec:paramamd}
Amdahl's law~\citep{amdahl} corresponds to the
special case of the USL equation (\ref{eqn:usl}) with $\beta = 0$.
Typically, it is used to quantify the achievable speedup:
\begin{equation}
C_A(N,\alpha) = \dfrac{N}{1 + \alpha (N-1)} \label{eqn:amdahl}
\end{equation}
for fine-grained parallelism.
Equation (\ref{eqn:amdahl}) is a rational function with $P(N)=N$ and 
$Q(N)=bN + c$; a linear polynomial with coefficient values \mbox{$a,b,c \geq 0$}.

Amdahl's law assumes that 
a single workload comprises a {\em parallel} portion and a remaining 
{\em serial} portion.
The serial portion or {\em serial fraction}, $0< \alpha <1$,
is the aggregate fraction of the workload that can only be 
executed sequentially on a single processor, i.e., the non-parallel portion. 

Another fundamental assumption is that the parallel portion  of the
workload can be partitioned into $N$ equal sub-tasks. If the size 
of these sub-tasks can be made progressively smaller, then the elapsed execution 
time will become dominated by the serial fraction such that 
$C(N) \sim 1/\alpha$ as $N \rightarrow \infty$. In other words, there is 
an asymptotic ceiling on achievable speedup shown as the horizontal line 
in Fig.~\ref{fig:funcmodels}.

\ssubsection{Gustafson's Linearization} \label{sec:paramgus}
Amdahl's law assumes the size of the work is fixed. Gustafson's
modification~\citep{gusto} is based on the idea of scaling up the size of
the work to match the $N$ processors~\citep[See][Chap. 4]{gcap}. 
This rescaling of the workload results in the
theoretical recovery of linear speedup
\begin{equation}
C_G(N,\alpha) = (1-\alpha) N  + \alpha \label{eqn:gusto}
\end{equation}
which is a rational function with $P(N)=bN + c$; a linear polynomial with 
coefficients \mbox{$a,b,c \geq 0$} and $Q(N)=1$, trivially. 

Unlike the USL, (\ref{eqn:gusto}) exhibits the peculiarity that
\mbox{$C_G(0,\alpha) = \alpha$}, i.e., there is non-zero capacity even
if there are no processors in the system! This artifact is shown as the blue line
near the origin in Fig.~\ref{fig:funcmodels} and must be regarded as
an unphysical side-effect of the Gustafson linearization of Amdahl's law.

Although $C_G(N,\alpha)$  has inspired various efforts for improving
parallel processing efficiencies, achieving truly linear speedup has
turned out to be extremely difficult in practice. Most recently,
(\ref{eqn:gusto}) has been proposed as a way to ``break Amdahl''
scalability for threaded applications running on multicore
processors~\citep{sutter}. Whether this claim will be effective, remains
to be seen. There is a significant literature of failed proposals for
defeating Amdahl's law~\citep[See, e.g.,][]{kleinrock,nelson,preparata}.
See Sect.~\ref{sec:zones} for some additional remarks based on the
results of this paper.

\ssubsection{Retrograde Scalability} \label{sec:retro}
The key difference between the USL and the other parametric models lies
in the fact that only the USL can successfully predict {\em retrograde}
scaling. See Fig.~\ref{fig:funcmodels}. In other words, if we think of
Gustafson's linear scalability as corresponding to ``equal bang for the
buck,'' and Amdahl's law as representing ``diminishing returns,'' then
the USL represents ``negative return on investment,'' or negative ROI.

Such negative ROI effects in application scalability are not the
exception but the norm. Figure~\ref{fig:USLwebsphere} shows an example
of WebSphere benchmark data fitted using {\em Mathematica}. The
retrograde effect is manifest. It is in this sense that the USL is 
considered to be {\em universal}.

\begin{thm}[Universality] \label{thm:universal}
The necessary and sufficient condition for the relative capacity $C(N)$
to be a {\em universal} scalability model is $P(N) = N$ and   
$Q(N)=aN^2 + bN + c$ with coefficients $a,b,c > 0$.
\end{thm}

\begin{prf}
The proof is best demonstrated by considering 
\uline{latency} rather than throughput. 
See~\citep{arxiv08} for details of this proof. \qed
\end{prf}

Interestingly, the proof establishes a similarity between the USL and
Brooks' law~\citep{brooks} for the management of software projects,
viz., {\em ``Adding more manpower to a late software project makes it
later.''} In this case, $N$ is interpreted as {\em people} rather than 
processes or processors.
Brooks' law is the analog of the negative ROI mentioned earlier.

Equation (\ref{eqn:usl}) clearly satisfies theorem~\ref{thm:universal}.
Next, we show that each of these parametric models has a more
fundamental meaning because they correspond to bounds on throughput for
a well-defined queueing model.

\ssection{QUEUEING MODELS} \label{sec:qmodels}
In Sects.~\ref{sec:prepman} and~\ref{sec:synq}, 
we develop some generalizations of the MRM that do  
not appear in the queueing-theory literature. 
First, we briefly review the standard MRM.

\begin{figure}[!htb]
\centering
\includegraphics[scale = 0.80]{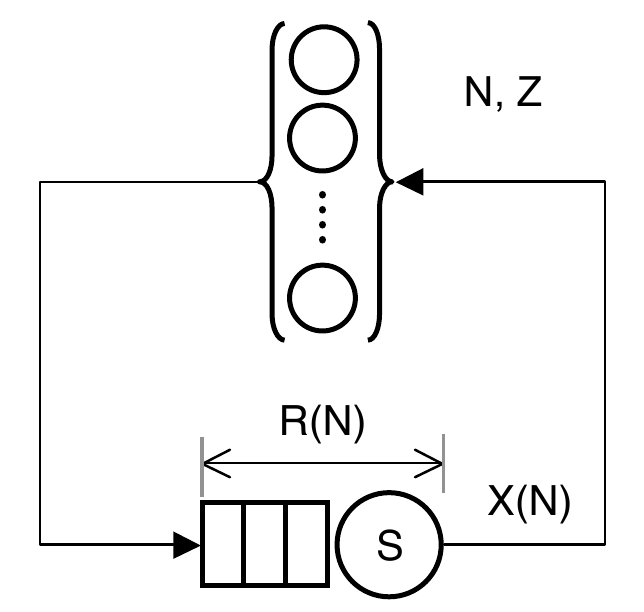} 
\caption{Machine repairman schematic} \label{fig:qrepairman}
\end{figure}

\ssubsection{Standard Repairman} \label{sec:repnormal}
The repairman queueing model~\citep{gross} is shown schematically in 
Fig.~\ref{fig:qrepairman}) and represents an assembly line comprising a finite
number of machines $p$ which break down after a mean lifetime $Z$.  A
repairman takes a mean time $S$ to repair each broken machine. If
multiple machines fail, the additional machines must queue for service
in FIFO order. The queue-theoretic notation for the MRM model, $M/M/1//N$,
implies exponentially distributed lifetimes and service periods with a
finite population $N$ of requests and buffering.

\begin{figure}[!htb]
\centering
\includegraphics[scale = 0.65]{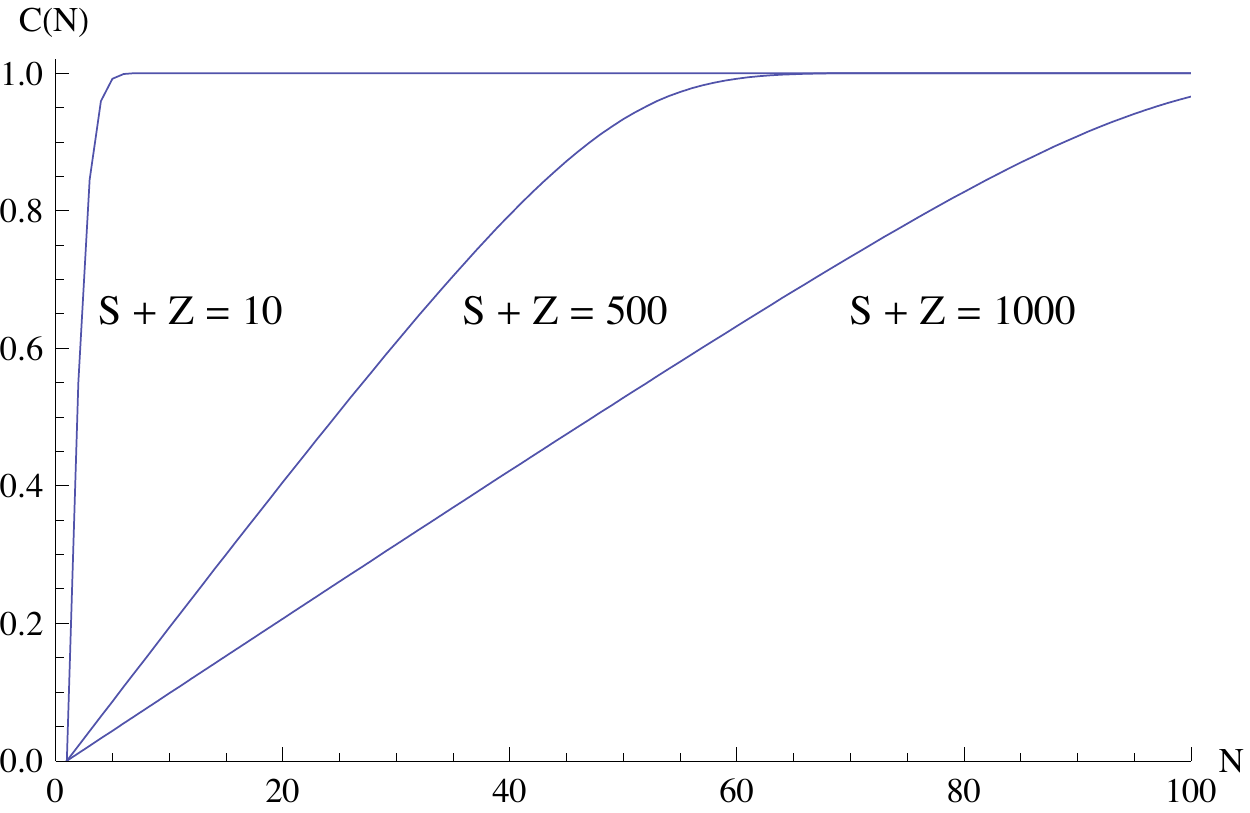} 
\caption{MRM throughput curves with normalized saturation values  
for round-trip times equal to 10, 500 and 1000 service units. No 
degradation is possible for any choice of $N$, $S$ or $Z$}
\label{fig:repxs}
\end{figure}

In steady state, $ZX$ machines are ``up'', while $Q$
are ``down'' for repairs, such that the total number of machines in
either state is given by \mbox{$N = Q + ZX$}. Rearranging this
expression we have:
\begin{equation}
Q = N - ZX
\end{equation}
and applying Little's law $Q=XR$ to
\begin{equation}
XR = N - ZX
\end{equation}
produces
\begin{equation}
R(N) = \dfrac{N}{X(N)} - Z   \label{eqn:mrmr}
\end{equation}
for the mean residence time at the repair station.
Rearranging (\ref{eqn:mrmr}) provides an expression for the mean MRM throughput   
as a function of $N$:
\begin{equation}
X(N) = \dfrac{N}{R(N) + Z} \label{eqn:mrmx}
\end{equation}
Several solutions of $X(N)/X(1)$ are shown in Fig.~\ref{fig:repxs}.

MRM has its origins in operations research associated with manufacturing
systems~\citep{gross}. However, because all queueing models are only
abstractions, MRM has found more widespread applications. The two most  
important applications for our discussion are:
\begin{description}\addtolength{\itemsep}{-0.5\baselineskip}
\item[Time-share systems:] 
$N$ represents users making requests to a
central processing facility. An MRM queueing model was applied to the
performance analysis of a research time-share computing system called
CTSS~\citep{scherr}; the precursor to UNIX.
\item[Multiprocessors:] 
$N$ represents CPUs or cores and the MRM 
represents the interconnect that allows the CPUs to communicate and
share data~\citep{balbo,reed}.
\end{description}

\begin{table}[!hbt]
\centering
\caption{Interpretation of the queueing metrics in Fig.~\ref{fig:qrepairman} 
where MRM: machine repairman, CMP: core multiprocessor and 
TSS: time-share system} \label{tab:mrm}
\begin{tabular}{cll}	
\hline
\bf Metric	& \bf Model	& \bf Interpretation \\
\hline
		& \sf MRM	& machines\\
$N$		& \sf CMP	& processors, cores\\		
		& \sf TSS	& processes, users\\
\hline
 		& \sf MRM	& up time\\ 			
$Z$		& \sf CMP	& execution period\\
		& \sf TSS	& think time\\
\hline
 		& \sf MRM	& service time\\	
$S$		& \sf CMP	& transmission time\\
		& \sf TSS 	& CPU time\\
\hline
	 	& \sf MRM	& residence time \\
$R(p)$	& \sf CMP	& interconnect latency\\
		& \sf TSS 	& queueing time\\
\hline
	 	& \sf MRM	& failure rate \\   	
$X(p)$	& \sf CMP	& bus bandwidth	\\			
		& \sf TSS 	& appln. throughput\\
\hline\end{tabular}
\end{table}

A summary of how to interpret the MRM queueing variables in each of the
described cases is provided in Table~\ref{tab:mrm}. The multiplicity of
MRM interpretations justifies the statement in
Sect.~\ref{sec:parametrics} that these parametric models can be applied
to both hardware and software scalability analysis. In particular, since the
USL model (\ref{eqn:usl}) does not presume any particular type of
application or topology, it can be applied equally well from multi-cores
to multi-tier systems. That detailed information is present in the USL,
but it is encoded in the two parameters: $\alpha$ and $\beta$.

Having reviewed the standard MRM, we new develop some generalizations that will 
be needed to make the connection between the MRM and the USL explicit.
These generalizations are: (i) state-dependent service times and (ii) synchronous queueing, 
which we treat in Sects.~\ref{sec:prepman} and~\ref{sec:synq}, respectively.

\ssubsection{Prepping Repairman} \label{sec:prepman}
Figure~\ref{fig:repxs} shows that the mean throughput is approximately
linear for values of $N$ near the origin (i.e., low load) and reaches a
saturation plateau at high loads when $N > S/(S+Z)$. Retrograde
throughput, of the type exhibited in Fig.~\ref{fig:USLwebsphere}, is not
possible in the standard MRM for any choice of $N$, $S$ or $Z$. However,
we do know that retrograde throughput is associated with load-dependent
servers in queueing models~\citep[See e.g.,][Chap. 10]{ppdq}.

The question then becomes: What should be the form of the
load-dependency such that the MRM produces {\em exactly} the retrograde
throughput exhibited by the USL? In principle, such load-dependence
could take any form. In Sects.~\ref{sec:synq} and~\ref{sec:simusl} we
show, rather surprisingly, that simple {\em linear} load-dependence is
required to produce USL behavior in the MRM.

Linear load-dependence in the context of the MRM means that the
repairman has to prepare up to $N$ failed machines in some way, e.g.,
rank them, {\em prior} to actually servicing them. In general, such
ranking will involve pairwise comparisons and this introduces an
additional delay that grows binomially with $N$, i.e.,
\mbox{$\binom{N}{2} = N (N-1)/2$}. Up to a factor of 2, this is
precisely the term in the denominator of (\ref{eqn:usl}). It is the
queueing analog of {\em negative ROI} discussed in
Sect.~\ref{sec:retro}.

It is important to realize that this additional delay due to 
preparations, is suffered by {\em all} of the enqueued machines before
the repairman commences service. For brevity, we shall hereafter refer
to this kind of load-dependent repairman as the {\em prepping
repairman}.

\ssubsection{Synchronous Queueing} \label{sec:synq}
One more condition is necessary in order to establish the connection
between the parametric models of Sect.~\ref{sec:parametrics} and the
MRM, viz., {\em synchronous queueing}. The reason for this requirement 
stems from the fact that the residence time $R(N)$ in
(\ref{eqn:mrmx}) can be quite arbitrary and, mathematically speaking, 
may not even possess an analytic form. But even if $R(N)$ does
have an analytic form, it is unlikely to be a polynomial in $N$ and
thus, will not produce rational functions like (\ref{eqn:ratfun}).

If, however, all machines were to break down {\em simultaneously}, the
queue length at the repairman would be maximized such that the residence
time becomes \mbox{$R(N)=NS$}, i.e., one machine in service and $(N-1)$
waiting. Synchronous queueing produces worst-case throughput and 
it therefore represents a lower bound~\citep{bounds} on (\ref{eqn:mrmx}):
\begin{equation}
\dfrac{N}{NS + Z} \leq X(N)  \label{eqn:lbound}
\end{equation}
In the context of multiprocessor scalability (see Table~\ref{tab:mrm}),
it is tantamount to all $N$ processors simultaneously exchanging data or
sending messages across the interconnect.

It is this synchronous queueing condition that causes the throughput
(\ref{eqn:mrmx}), in both the standard and load-dependent MRM models, to
conform to (\ref{eqn:ratfun}) and thus provide the connection with the
rational functions discussed in Sect.~\ref{sec:parametrics}.

\begin{thm}[Main theorem] \label{thm:usl} 
The universal scalability law (\ref{eqn:usl}) is equivalent to the 
synchronous bound on relative capacity in the MRM with
linear load-dependent service rate.
\end{thm}

\begin{prf}[Sketch]
Under synchronous queueing in MRM, when the first request is in 
service the mean waiting time for the remainder is given by 
\begin{equation}
W = (Q-1) \, S			\label{eqn:sync}
\end{equation} 
where $Q$ is the mean number of requests in the system. 
Now, let the service time be load-dependent such that:
\begin{equation*}
S(Q) = c \, Q S
\end{equation*}
with $c$ a constant of proportionality. For synchronous queueing 
\mbox{$Q \equiv N$}, so we can rewrite (\ref{eqn:sync}) as:
\begin{equation}
W = c \, (Q-1) \,  Q S = c \, N \, (N-1)  \, S \label{eqn:quadp}
\end{equation} 
Expressed as relative relative throughput, (\ref{eqn:quadp}) appears in the
denominator of (\ref{eqn:usl}) as the $N(N-1)$ term. 
The detailed proof appears in~\citep{arxiv08}. \qed
\end{prf}

If we consider synchronous queueing in the standard MRM, i.e., without
load-dependent service, we recover Amdahl's law, which is also the
special case of USL with $\beta = 0$ in (\ref{eqn:usl}).

\begin{cor} \label{cor:amdmrm} 
Amdahl's law (\ref{eqn:amdahl}) is equivalent to the relative throughput
due to synchronous queueing in the standard MRM with
mean execution time $Z$ and constant service rate $S$.
\end{cor}

\begin{prf}
The proof  requires the identity 
\begin{equation}
\alpha = \frac{S}{ S + Z } \, . \label{eqn:serialfract}
\end{equation}
See~\citep[Appendix A]{gcap} and~\citep{arxiv02} for details. \qed
\end{prf}

\begin{cor} \label{cor:gustomrm} 
Gustafson's law (\ref{eqn:gusto}) corresponds to the rescaling 
\mbox{$Z \mapsto pZ$} in the MRM. 
\end{cor}

\begin{prf}
See~\citep{arxiv08} as well as the simulation results in 
Sect.~\ref{sec:simgus}.
\end{prf}

The precise nature of the synchronization discussed here turns out to be
rather subtle. To see this, consider the case where all $N$ machines
have the same deterministic $Z$ period. At the end of the first $Z$
period, all $N$ machines will enqueue at the MRM simultaneously.
By definition, however, the machines are serviced serially, so they will
return to the the operational phase  (parallel execution in the top
portion of Fig.~\ref{fig:qrepairman}) separately and thereafter will always
return to the repairman at different times. In other words, even if the
queueing system is started with synchronized visits to the repairman,
that synchronization is immediately lost after the first tour because it
is destroyed by the very process of queueing.

Unfortunately, the analytic equations used in this section only provide
a steady-state view of the MRM, so we cannot discern the details of the
stable synchronization process. Consequently, we turn to
discrete-event simulation models~\citep{holtsim} as an {\em exploratory tool}, rather
than a predictive tool. As we shall see in Sect.~\ref{sec:zones}, the
synchronous MRM simulations also provide deeper insight into potential
performance tuning opportunities in real systems.

\begin{figure}[!htb]
\centering
\includegraphics[scale = 0.55]{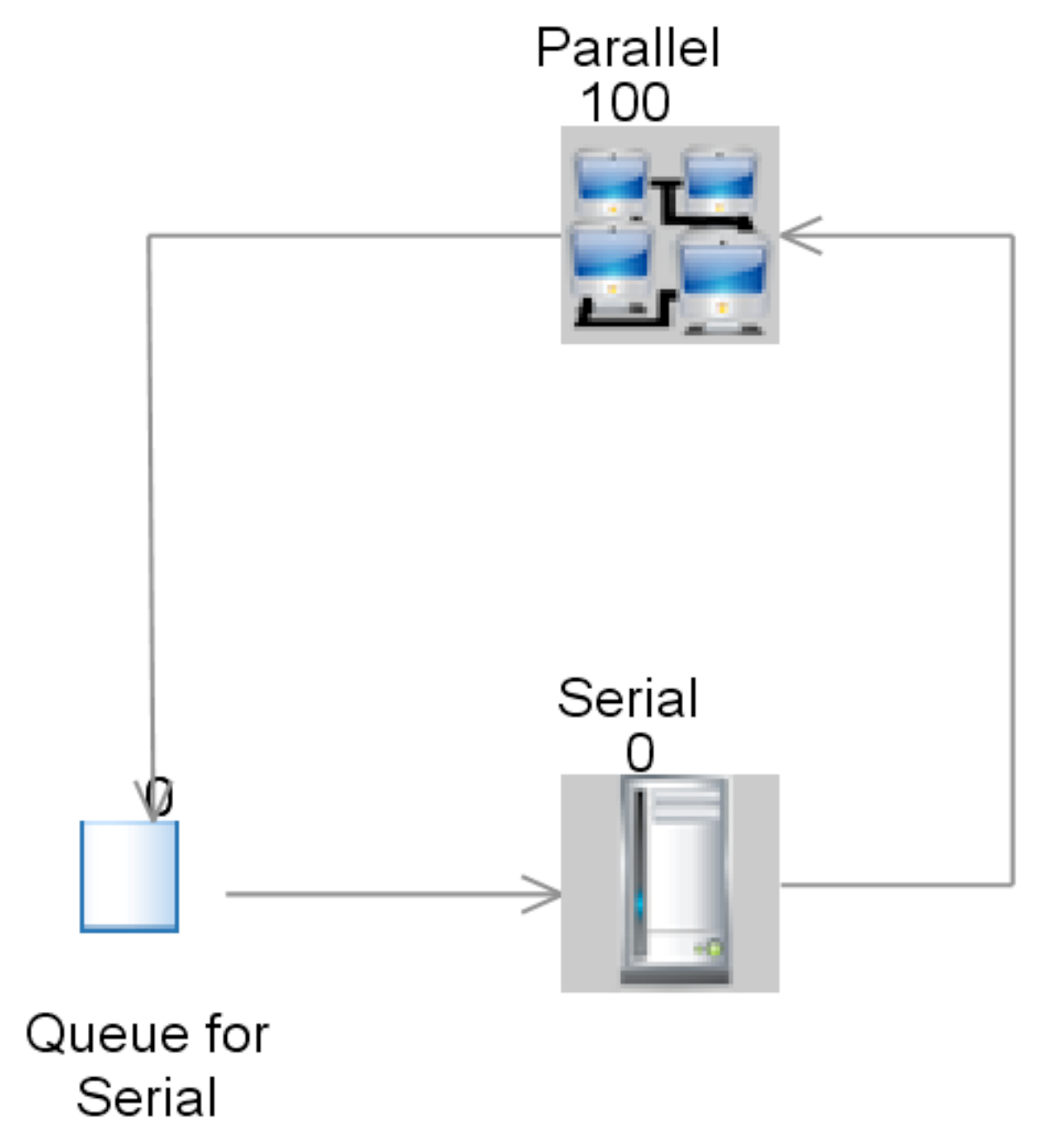} 
\caption{SIMUL8 model corresponding to Fig.~\ref{fig:qrepairman}}
\label{fig:simul8}
\end{figure}

\ssection{SIMULATION MODELS} \label{sec:sims}
From a basic modeling standpoint, using discrete event
simulation, these look relatively the same with a fixed
number of requests circulating through a closed system with a {\em wait time} 
and {\em service time} associated with them.  The basic model is shown in 
Fig.~\ref{fig:simul8}.

There is a place where the requests are running in parallel on
multiple CPUs, and another place where the requests are serialized on a
single CPU.  The cases differ in how requests make the transitions
between the two modes of operation.

The simulation models are intended to clarify how the analytical results
in Sects.~\ref{sec:parametrics} and~\ref{sec:qmodels} relate to the way
an actual application executes.  (cf. Fig.~\ref{fig:USLwebsphere}) They
will be used to demonstrate what features of an architecture might drive
an application to the various performance regions or zones defined in
Fig.~\ref{fig:zones}. Some of us might be mathematically challenged and
prefer to see something running on a real platform, on which we can
measure performance, that exhibit the characteristics defined by the
analytical equations. In lieu of a real platform, we use simulation
models. The criteria used in each of the models on how the parameters
are used and how they related to the real world will be defined.
Hopefully this will provide the reader with a better understanding of
how each of these models are representative of situations that they
might have seen on their own computer systems.

We have presented a number of mathematical equations that
represent the expected throughput of various models of computer
scalability.  There are four models discussed in this section for which a
discrete simulation queueing model will be created to show that there is
a correlation between what the analytic models predict and the results of
the simulation models (and thus real platforms). 

Discrete event simulation can be defined as ``the operation of a system
is represented as a chronological sequence of events.  A common exercise
in learning how to build discrete-event simulations is to model a queue,
such as customers arriving at a bank to be served by a teller. In this
example, the system entities are CUSTOMER-QUEUE and TELLERS.''   As
mentioned in Sect.~\ref{sec:intro}, this approach has been used to
create models for computer performance evaluation.  Simulation
represents a real system by modeling the important characteristics.  For
the models in this paper, there are three objects that are used in the
modeling:
\begin{enumerate}\addtolength{\itemsep}{-0.5\baselineskip}
\item {\bf Resource} to be consumed (CPU cycles in this case)
\item {\bf Consumer} of the resource (CPU actively working on a request)
\item {\bf Queue} to hold requests for the CPU if it is not available
\end{enumerate}

A word of clarification might be helpful in light of the comment made in
Sect.~\ref{sec:intro}, regarding the common impass of needing to
parameterize queueing models with service times that are often not
available in standard performance measurements. 
We are not using simulation models to make performance predictions in the 
usual sense.  The simulation models
discussed in this paper are constructed to explore the underlying
dynamics of the analytic scalability models in
Sect.~\ref{sec:parametrics} and~\ref{sec:qmodels}. In order to reveal
the connection between the simulation models and the analytic models, it
turns out that we only need to define the {\em ratios} between the
queueing variables $N$, $S$ and $Z$ in Table~\ref{tab:mrm}. The actual
numbers can be any numeric values, e.g., $S = 1.0$ second. In this
sense, we are free to construct our simulation models because they do not 
require measured service times.

\ssubsection{Repairman Simulation} \label{sec:simman}
The first simulation model we consider is the standard MRM 
defined in Sect.~\ref{sec:repnormal}.  This model
is a closed  queueing model and can be used to explain the performance
of a system where jobs can run in parallel (no contention for the CPU
resource), but every so often they enter a serial activity where only
one request can be processed at a time (e.g., programs requesting a lock
on a common table).  In this case multiple requests may queue up waiting
to progress through the serial portion.  The other characteristic of the
MRM  is that it is also defined as an asynchronous model in
that the jobs may enter the serial portion independent of one another.

This model (Fig.~\ref{fig:simul8}) was constructed using the SIMUL8
event-based simulator.  Here the flow of work in closed system is
represented.   As soon as the serial portion is completed, the request
can start execution in parallel with other requests.  The ``Serial'' (S)
and ``Parallel'' (P) objects are consumers of the ``CPU'' resource,
which in this case has 100 CPUs available so that all 100 requests can
execute at the same time in the Parallel object.  The Serial object has
the restriction that only one CPU can be used at a time, forcing
outstanding requests to be queued and processed sequentially.

Besides the number of CPUs, the other important parameter in the model
in the percent of CPU time spent in the serial execution as compared to
the parallel execution.  For all the models in this paper, 10\% of the
time is serial operation.  If the time to process a unit of work is 1 
second, then $0.9$ seconds are spent in the parallel path and 0.1 seconds
in the serial path.  The bottleneck of the system is the serial path and
will limit the throughput to 10 jobs per second ($1/0.1$).  In the model
the timings in the Parallel and Serial objects are an exponential
distribution based on the service time given above.

\begin{figure}[!htb]
\centering
\includegraphics[scale = 0.55]{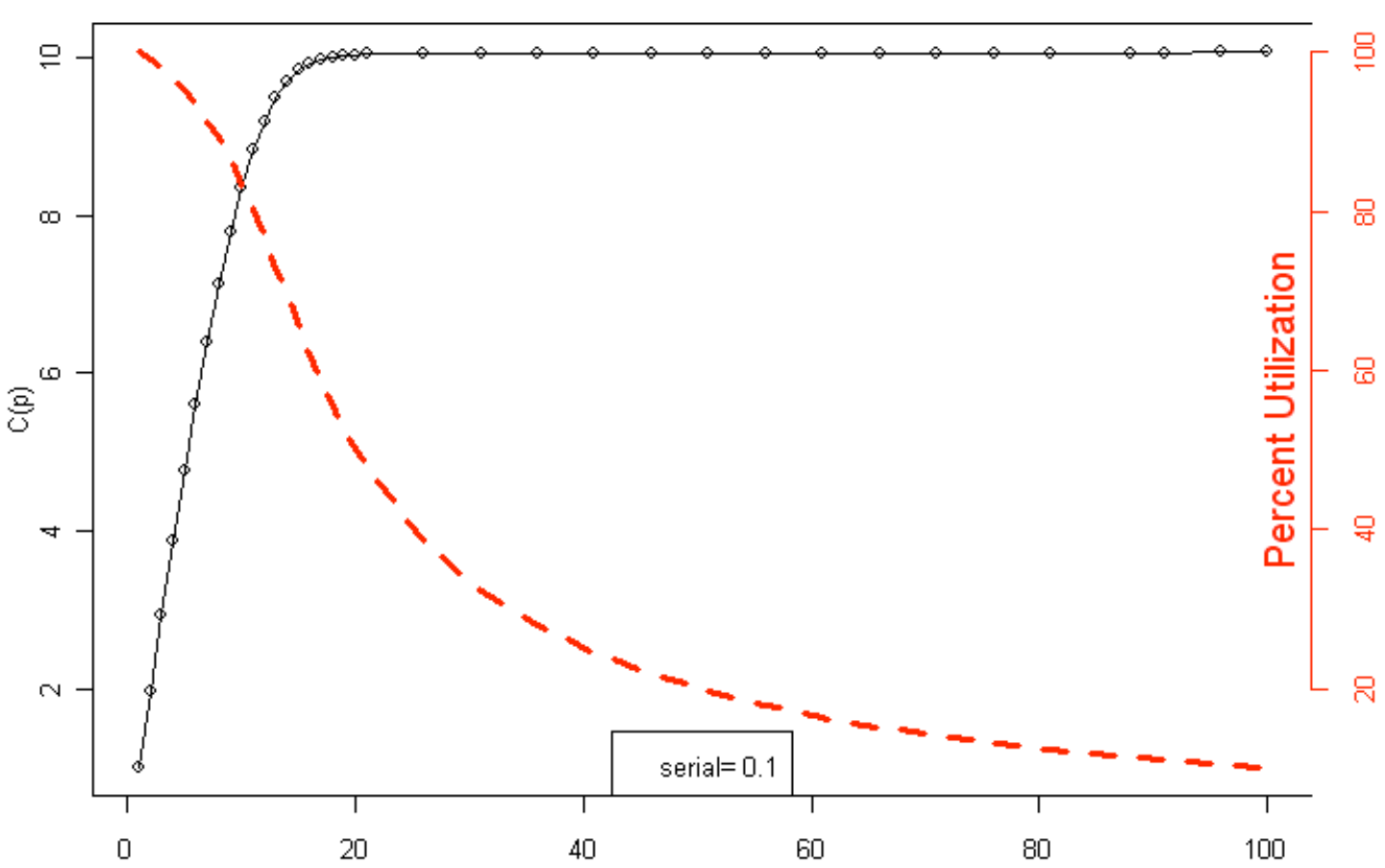} 
\caption{Throughput of the conventional MRM}
\label{fig:simrepx}
\end{figure}

So what we want to do in the model is to run a series of configurations
and measure the throughput of the system.  In this case the percentage
in the serial path is fixed and the number of initial requests is
varied.  The number of available CPUs is set to the number of initial
requests.  The throughput will be normalized to a system with a single
request.  Plotting this curve will provide an indication of the effect
of adding more CPUs to handle the load as the number of requests in the
system also increases.  Fig.~\ref{fig:simrepx} is the result of the
model using the percentages above.

The x-axis presents the number of CPUs (or requests) in the system.  As
you can see, we reach a normalized throughput of 10 relatively quickly. 
Plotted on the secondary y-axis on the right is the system utilization
as would be measured by the operating system.  Notice that when we reach
100 CPUs, the overall system utilization is only 10\%, or as indicated by
the graph, only 10 CPUs are doing useful work; the other are idle since
the tasks that were executing on them are in the queue waiting for
serial execution.

From Table~\ref{tab:mrm}, we know that the MRM is representative of a
multi-user ($N$), client/server, system where asynchronous requests are
contending for a common resource. To minimize loss of 
throughput, the serial portion must be reduced or the work partitioned
so the CPUs can be used more effectively.

The model can provide the direct numbers in the case of the MRM 
(e.g., an average of $89.7$ requests in the queue and a response time of
$9.07$ seconds for a request to make it through the serial portion). 
But we will see in the other models we will not be able to directly
measure these values because of the way the model handles the
synchronization of the Parallel and Serial operation.  But we can derive
these values by keeping track of the amount of CPU consumed in the
Parallel and Serial components.

For the case of the MRM with $N=100$ requests, the model was run for
$3,000$ seconds, and $27,206$ CPU seconds were consumed in Parallel and
$2,999$ in the Serial object.  There were $30,118$ requests processed in this
time.  We can also determine the average number of requests by
evaluating the number of CPU sec/sec consumed, which will indicate the
number of CPUs required to handle the requests.
\begin{align*}
N_P &= \dfrac{27206~\text{CPU}}{3000~\text{sec}} = 9.07~~\text{CPU sec/sec}\\
N_S &= \dfrac{2999~\text{CPU}}{3000~\text{sec}} = 1.00~~\text{CPU sec/sec}
\end{align*}
We have the number of requests in the Parallel and Serial objects, so
therefore the remainder must be in the queue for the Serial object:
\begin{equation*}
N_Q = 100-9.07-1.00 = 89.93
\end{equation*}
The response time can be computed using Little's law:
\begin{equation*}
R = \dfrac{N_Q + N_S}{X} = \dfrac{89.93 + 1.00}{30118/3000} = 9.06~\text{sec} 
\end{equation*}
where $X$ is the throughput of the MRM.

\ssubsection{Synchronization Gate} \label{sec:syncgate}
The difference between the MRM and Amdahl models is the way that work is
returned to the parallel/think processing node. In standard MRM, you can
imagine the repairman being presented with a bin of parts that need so
have service done on them before returning to operation. The repairman
will take a part out of the ``bin,'' do something to it for time $S$,
and then return the part back into operation before picking out the next
one from the bin.

\begin{figure}[!htb]
\centering
\includegraphics[scale = 0.45]{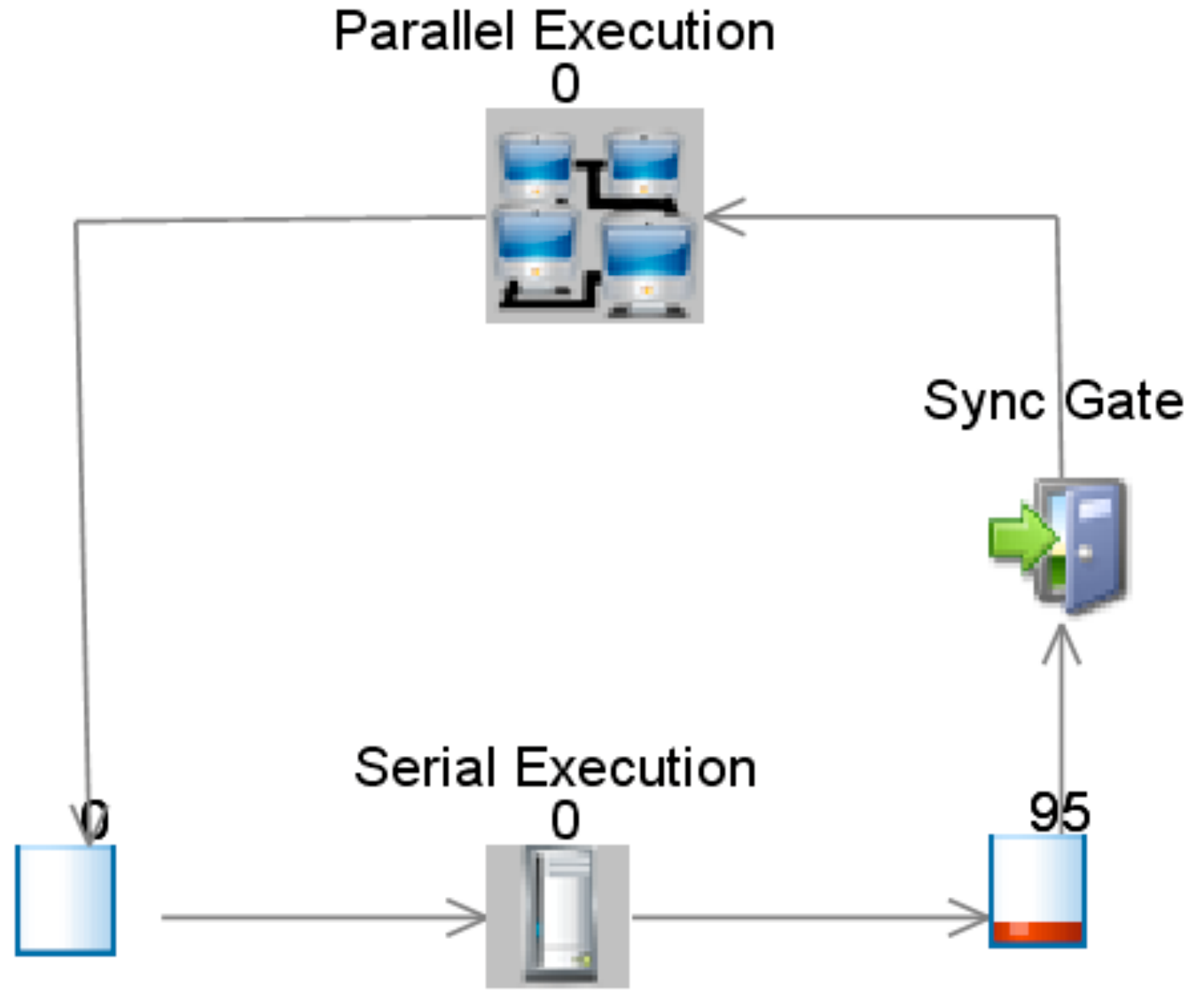} 
\caption{MRM with synchronization gate}
\label{fig:syncgate}
\end{figure}

In the case of Amdahl, the repairman will still receive a bin of parts
to service, but instead of immediately returning the part to operation
after it has been serviced, the repairman put the part in an output bin
and only when the input bin has been completely serviced are all the
parts returned to operation. As shown in Fig.~\ref{fig:syncgate}, there is
effectively a ``gate'' that prevents the release of  parts until the
repairman has repaired all of them.

Only after all the jobs were in the right buffer after the serial
execution were they released back to the parallel execution. But this
had some problems in that in the real world jobs do not necessarily all
stop at the same time and request some serial operation. There were also
some limitations on the types of distributions that could be used for
the serial and parallel timings. To overcome these limitations, consider
a time sharing system where is one of the jobs needs to run on the
serial path, it will stop or suspend all the jobs on the parallel path
until it completes. That is effectively what the gate is doing in the
figure above. The only impact on the parallel jobs is that their elapsed
time has been increased by the time it took the serial job to execute.
With this approach, jobs can come out of the parallel path in a random
fashion and the overall effect is the same as using a gate.

\begin{figure}[!htb]
\centering
\includegraphics[scale = 0.75]{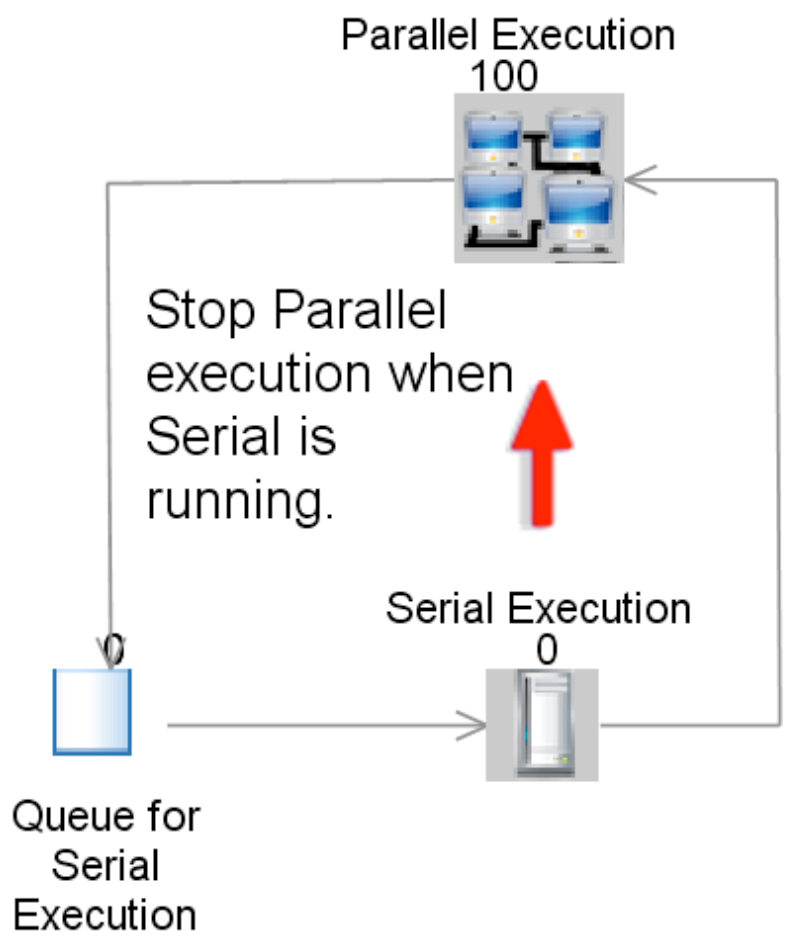} 
\caption{Amdahl model in SIMUL8}
\label{fig:simamd}
\end{figure}

So the new model of the system looks like Fig.~\ref{fig:simamd} where as soon as a
job starts execution on the serial path, a suspend signal is sent to the
parallel portion (could be on the next clock tick all the jobs are put
in a suspended mode). The serial job will execute and when it is
complete, the suspended jobs will be restarted and the job that
completed the serial path is returned to execute in the parallel path.

\ssubsection{Amdahl Simulation} \label{sec:simamd}
Figure~\ref{fig:simamd} is very similar to the MRM
(Fig.~\ref{fig:simul8}) with the exception that if there is a request
running in the Serial object, the processing in the Parallel object will
be suspended for the duration of the Serial execution.  This is
equivalent to all the requests trying to get at the same resource at the
same time and each one is serialized and none can restart until all the
serial requests have been handled.
 
In the simulation model, when a request starts to operate in the Serial
object, a ``breakdown'' will be generated for the Parallel object.  All
work will stop (be suspended) in the Parallel object for the duration of
execution on the Serial object and will then pick up where they left
off.  This effectively extends the elapsed time in Parallel, but the CPU
time stays the same.  This is contrasted with the MRM,  where
the elapsed time was the same as the CPU time in the Parallel object.

\begin{figure}[!htb]
\centering
\includegraphics[scale = 0.55]{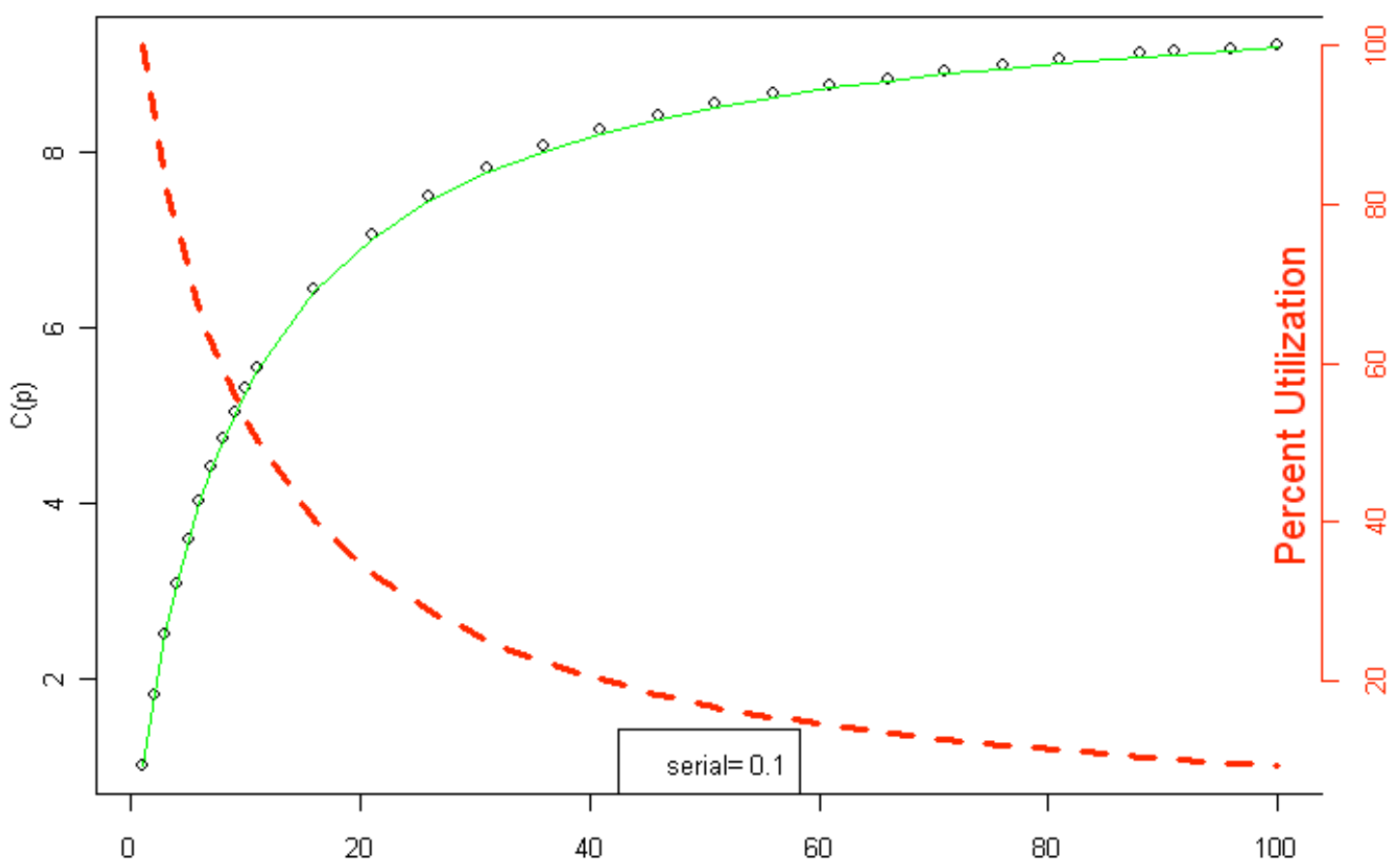} 
\caption{Amdahl throughput}
\label{fig:simamdx}
\end{figure}

In Fig~\ref{fig:simamdx}, the points represent the output from the model and the
green line represents the normalized throughput (speedup) as predicted
by Amdahl's law.  The model, which represents a ``synchronous'' MRM, 
correlates exactly with the predicted results from Amdahl's law. 
The limit in this case is also a speedup of $10$.

The reason for calculating the number of requests in each object, and
the response time, in the following manner is that if you look in the
model, the number of requests in the queue averages zero.  This is
because as soon as a request is completed in the Parallel object, it is
sent to the Serial object where it immediately begins execution, since a
request can only complete in the Parallel object if it is not suspended
which mean there is no request being processed on the Serial object.

This model was run for $3000$ seconds and had a throughput of $27624$ 
requests.  The number of requests in each of the objects and the
response times are:
\begin{align*}
N_P &= \dfrac{24842~\text{CPU}}{3000~\text{sec}} = 8.28~~\text{CPU sec/sec}\\
N_S &= \dfrac{2750~\text{CPU}}{3000~\text{sec}}  = 0.92~~\text{CPU sec/sec}\\
N_Q &= 100-8.28 - 0.92  = 90.80\\
R   &= \dfrac{90.80 + 0.92}{27624/3000}  = 9.96~\text{sec}
\end{align*}

\begin{figure}[!htb]
\centering
\includegraphics[scale = 0.65]{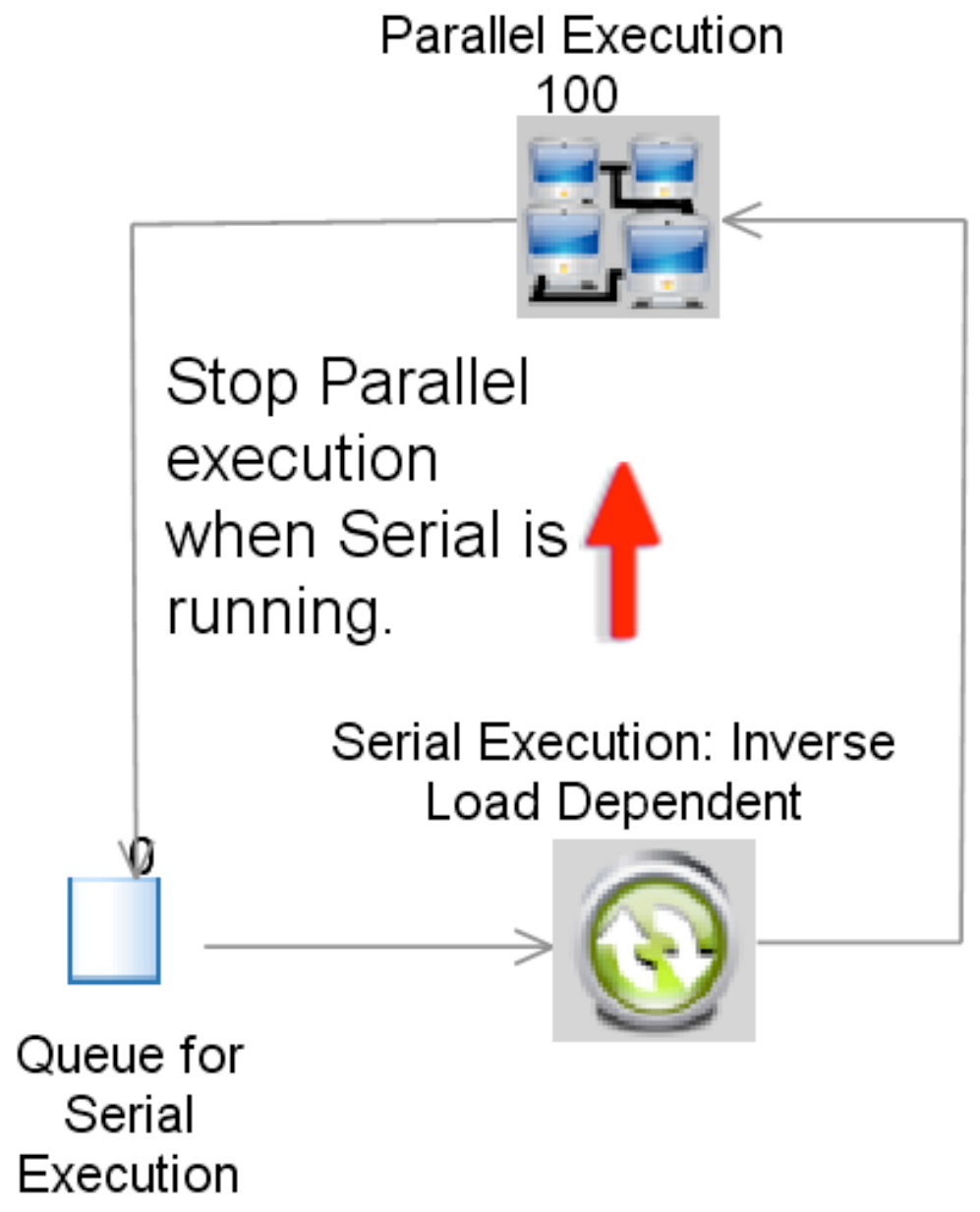} 
\caption{Gustafson model in SIMUL8}
\label{fig:simgus}
\end{figure}

\ssubsection{Gustafson Simulation} \label{sec:simgus}
The next simulation model represents Gustafson's parametric equation
discussed in Sect.~\ref{sec:paramgus}. This simulation model
(Fig.~\ref{fig:simgus}) is similar to the Amdahl model in that requests
are running in parallel until they all need the serial portion at the
same time, but instead of each request being processed sequentially
through the serial portion, all the requests are processed as a batch on
a single CPU with a constant service time.  This might represent locking
on a common table that takes the same time to update, no matter how many
requests are waiting.  In the Amdahl model, each request must be
processed sequentially, so that when there are more requests, there is
an additional increase in the time spent in the serial portion.  For the
Gustafson model, the time spent in the serial portion is constant and is
restricted to using a single CPU.

\begin{figure}[!htb]
\centering
\includegraphics[scale = 0.5]{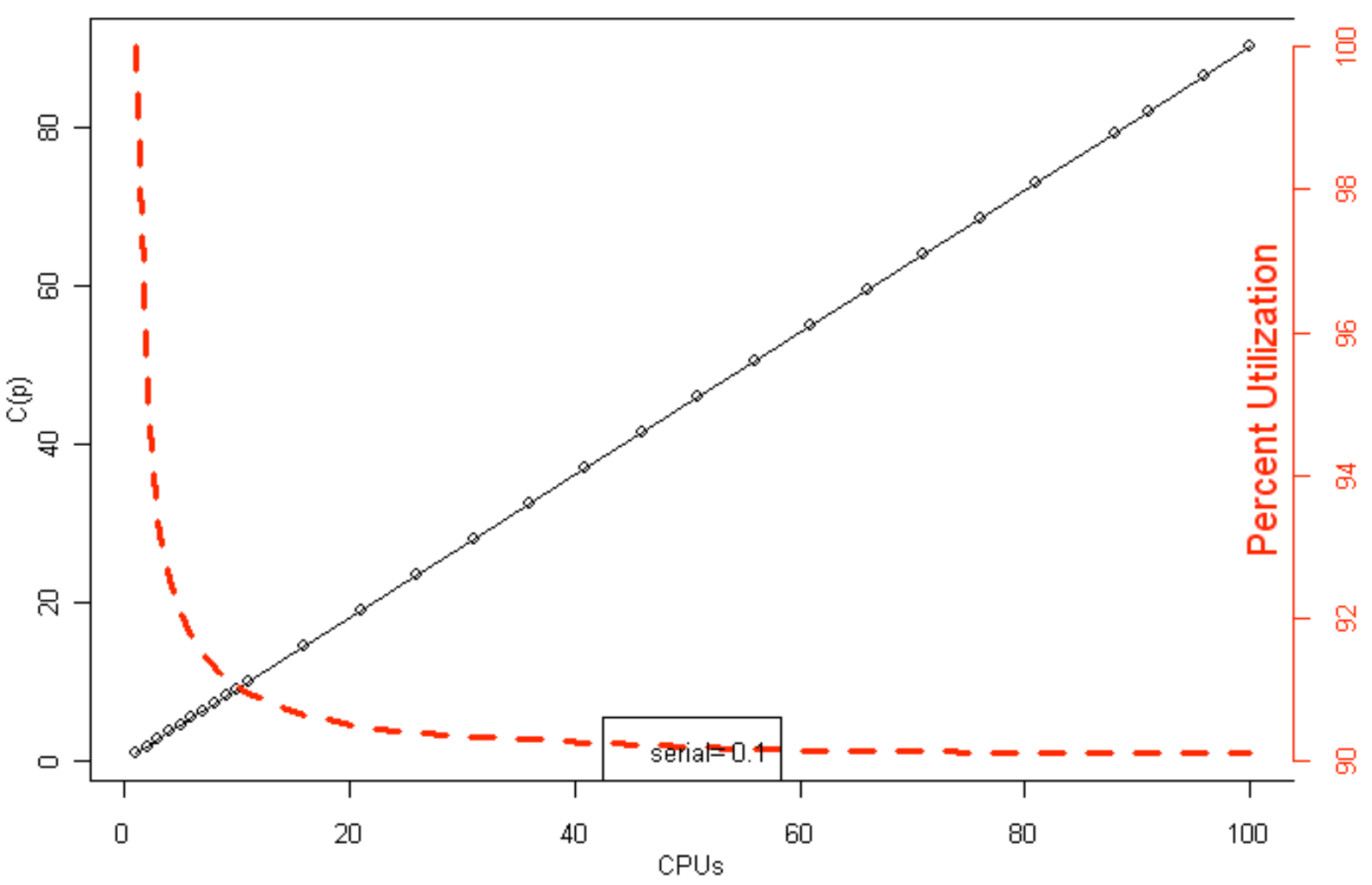} 
\caption{Gustafson throughput}
\label{fig:simgusx}
\end{figure}

Refering to Fig.~\ref{fig:simgusx}, the speedup is linear (diagonal
line). The system utilization (curve) falls from 100\% and approaches an
asymptote at 90\% as $N$ increases. As mentioned in the last paragraph
of Sect.~\ref{sec:sims}, this follows from the ratio choice of 10\% for
the serial fraction defined in (\ref{eqn:serialfract}). The parallel
timing is a fixed/constant distribution so that all the requests ask for
the serial portion at the same time and the serial timing is
exponential.

Gustafson's law is a linear function; the blue line in
Fig.~\ref{fig:funcmodels}. Fitting a linear regression
model~\citep{holtR} of the form \mbox{$C(N)= m N + c$} produces 
\mbox{$C(N)= 0.909 N + 0.084$} with the coefficient of
determination $R^2=1$, indicating a nearly perfect fit to the simulation
data. To within experimental error the gradient $m=0.909$ represents the
gradient in Gustafson's law, i.e., $(1-\alpha) = 0.90$, since we chose
$\alpha = 0.10$ in all our simulations. Similarly, rounding the
intercept value to one significant digit gives $c \simeq 0.1$, in
agreement with $\alpha = 0.10$ as the intercept in (\ref{eqn:gusto}).

This model was run for $3000$ seconds and had a throughput of $299608$ 
requests. 
\begin{align*}
N_P &= \dfrac{270033~\text{CPU}}{3000~\text{sec}} 	= 90.0~~\text{CPU sec/sec}\\
N_S &= \dfrac{299~\text{CPU}}{3000~\text{sec}}  	= 0.10~~\text{CPU sec/sec}\\
N_Q &= 100-90.0 - 0.10  							= 9.9\\
R   &= \dfrac{9.9 + 0.10}{299608/3000}  			= 0.10~\text{sec} 
\end{align*}
The response time remains the same as the service time ($0.1$ seconds),
which is what is expected with the Gustafson model.

\ssubsection{USL Simulation} \label{sec:simusl}
The next case is the USL model defined in Sect.~\ref{sec:paramusl}. 
This is similar to
the Amdahl model, but with the addition of a load-dependent serial
server.  What this means is that as the number of requests waiting for
serial service increases, the amount of time that it take the serial
portion to run also increases.  This is equivalent to a program that
might read through the waiting queue to pick the highest priority
request to process.  So each time a message is retrieved from the queue,
the entire queue is searched.  This might not take a lot
of time, but as the number of outstanding requests increases, so does
the serial processing time.
 
\begin{figure}[!ht]
\centering
\includegraphics[scale = 0.5]{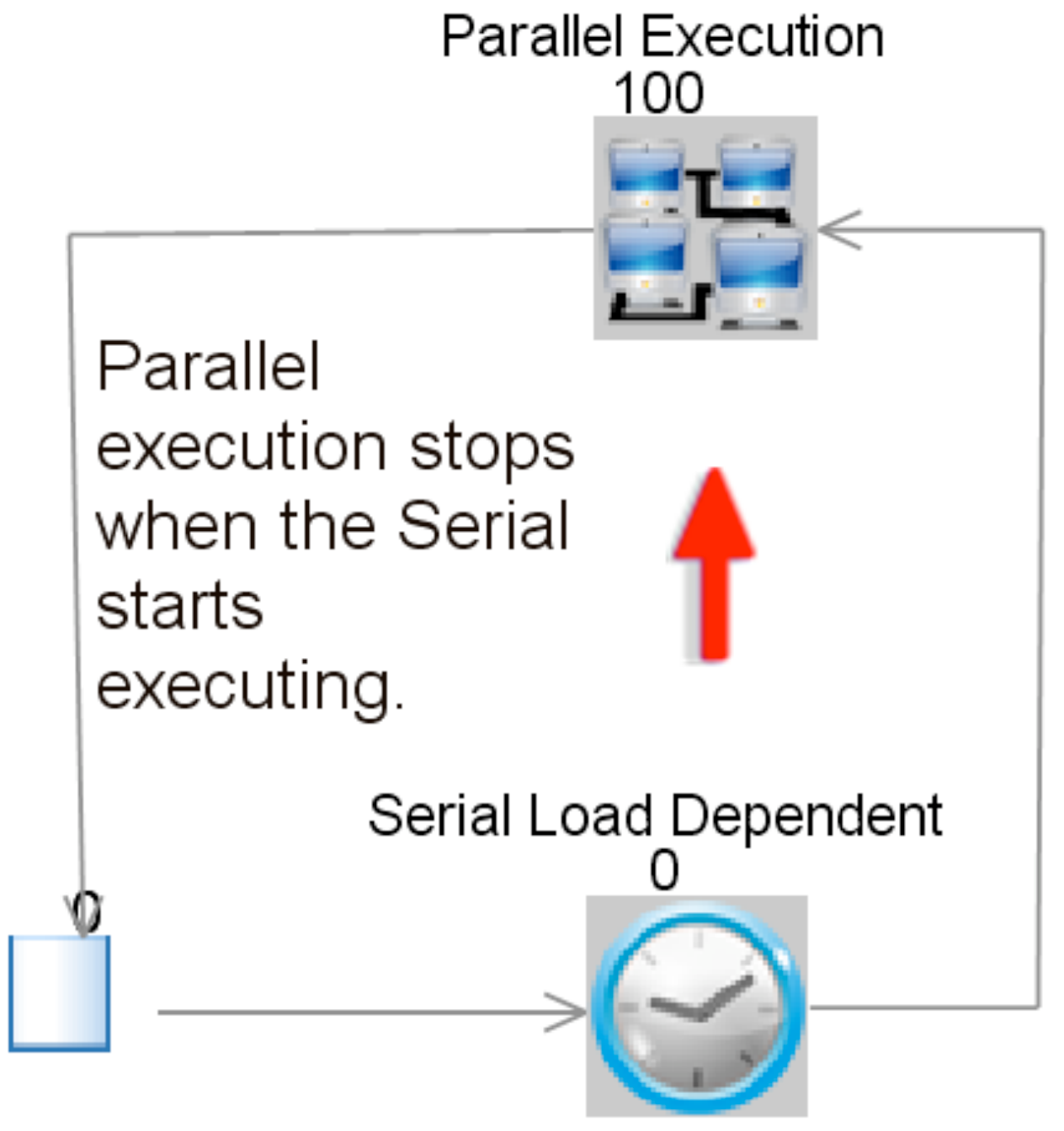} 
\caption{USL model in SIMUL8}
\label{fig:simusl}
\end{figure} 

This model was run with 10\% of the time spent in the serial portion and
an additional 0.1\% increase in the serial service time for every request
that is waiting in the queue.

\begin{figure}[!hb]
\centering
\includegraphics[scale = 0.6]{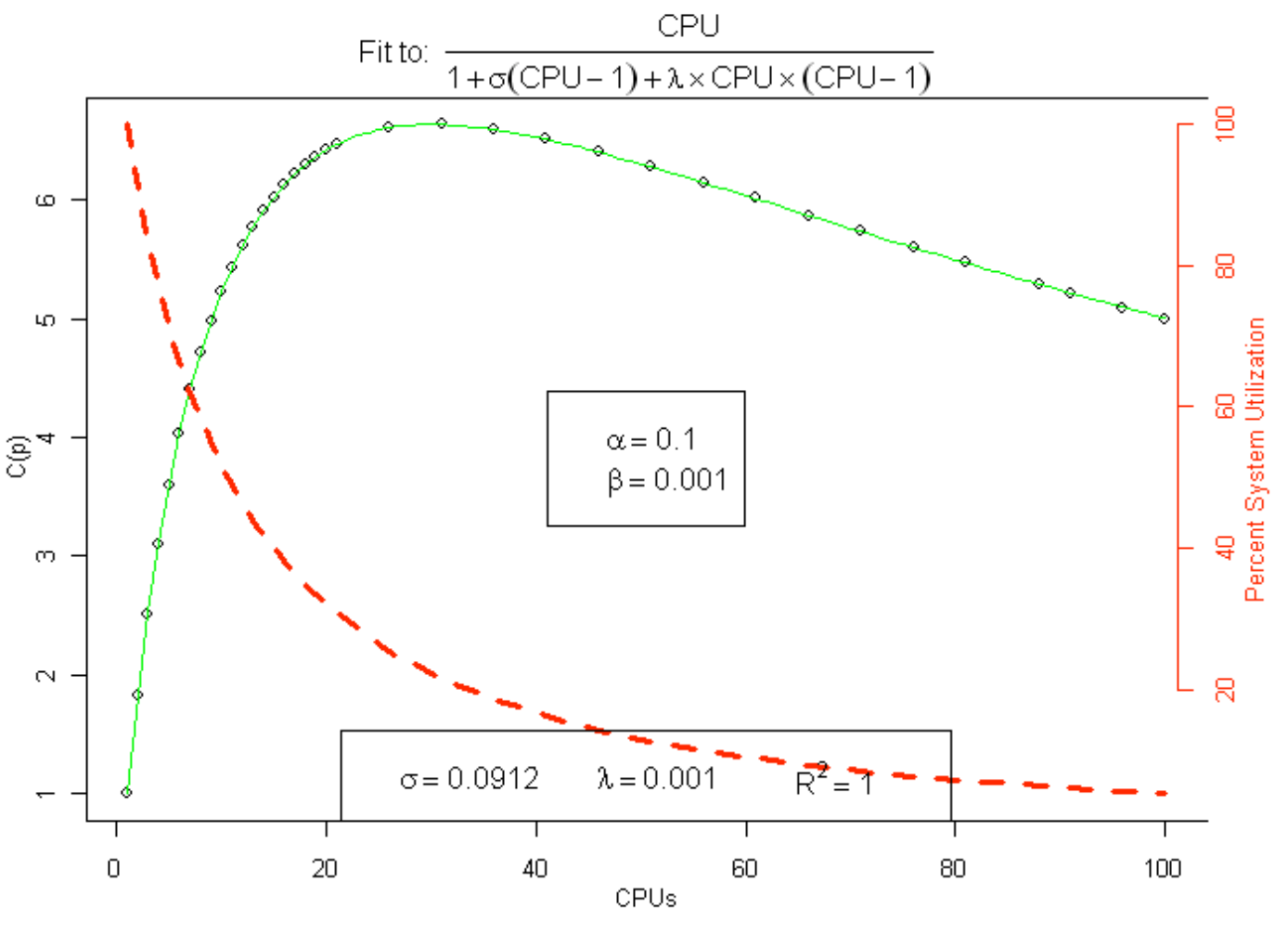} 
\caption{USL throughput}
\label{fig:simuslx}
\end{figure} 

Figure~\ref{fig:simuslx} illustrates a retrograde speedup due to the
increase in processing time based on the number of requests outstanding
in the queue.  The points on the graph are the output from the
simulation model.  The green line is a fit to the Universal Scalability
equation. The legend at the bottom of the graph is output from fitting
the data points to the equation. The results from the model match the
equation ($R^2=1$), indicating that Universal Scalability is the
equivalent of an Amdahl model, with the addition of a load-dependent
serial server.

The model was run for $3000$ seconds and had a throughput of $14727$ requests.
\begin{align*}
N_P &= \dfrac{13245~\text{CPU}}{3000~\text{sec}} = 8.28~~\text{CPU sec/sec}\\
N_S &= \dfrac{2867~\text{CPU}}{3000~\text{sec}} = 0.92~~\text{CPU sec/sec}\\
N_Q &= 100-4.42 - 0.96 = 94.62\\
R   &= \dfrac{94.62 + 0.96}{14727/3000} = 19.47~\text{sec} 
\end{align*}

\ssubsection{Generalized Distributions} \label{sec:anydist}
For the simulations in Sect.~\ref{sec:sims}, exponential distributions were used in
all cases since they are the basis for solving most analytical equations.
See Sect.~\ref{sec:repnormal}.
But we can show with the simulation models that any distributions
can be chosen for the service times, thereby extending the applicability of
the USL.

\begin{figure}[!htb]
\centering
\includegraphics[scale = 0.6]{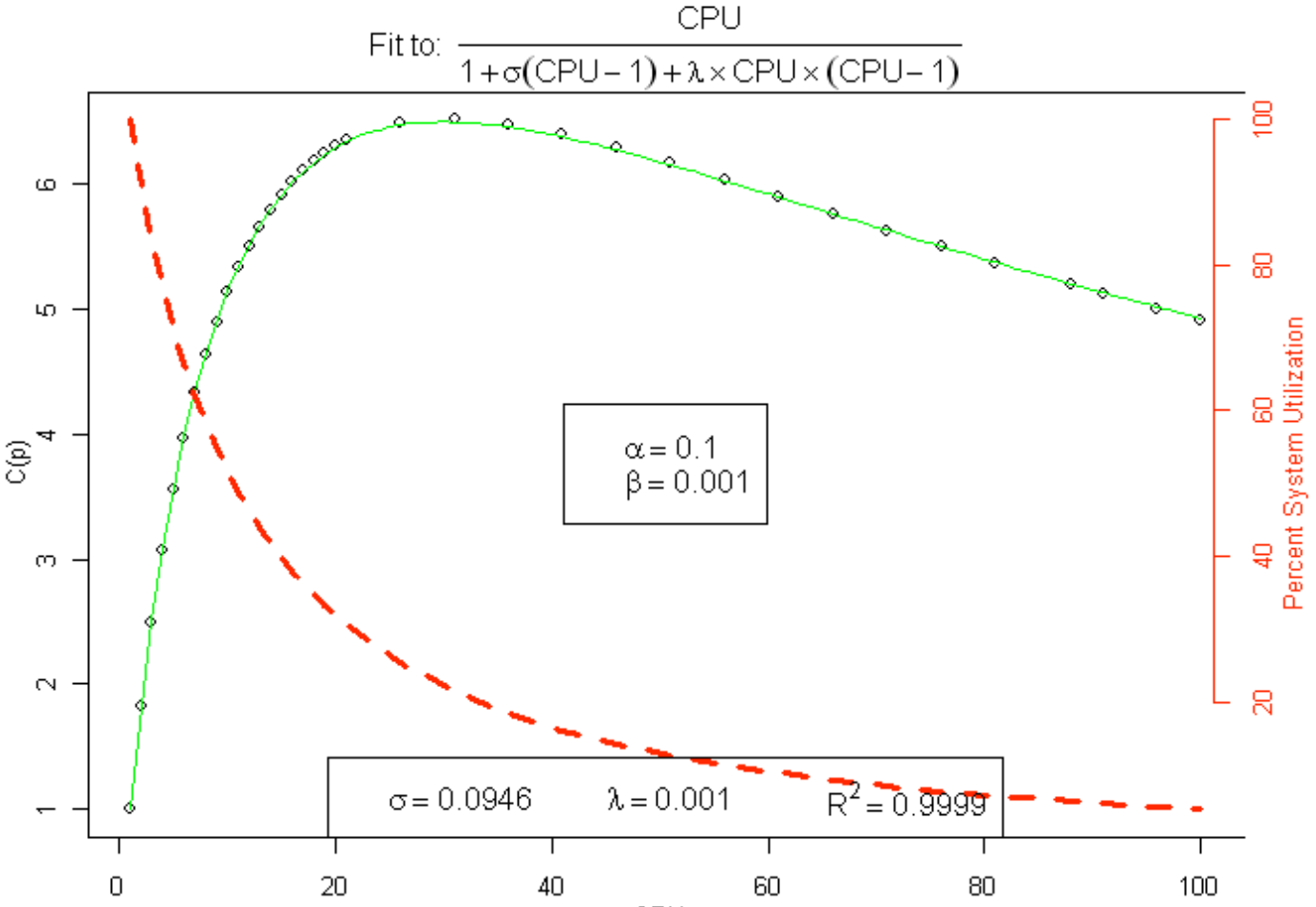} 
\caption{USL model with generalized distributions}
\label{fig:figure9}
\end{figure}

\begin{figure}[!htb]
\centering
\includegraphics[scale = 0.70]{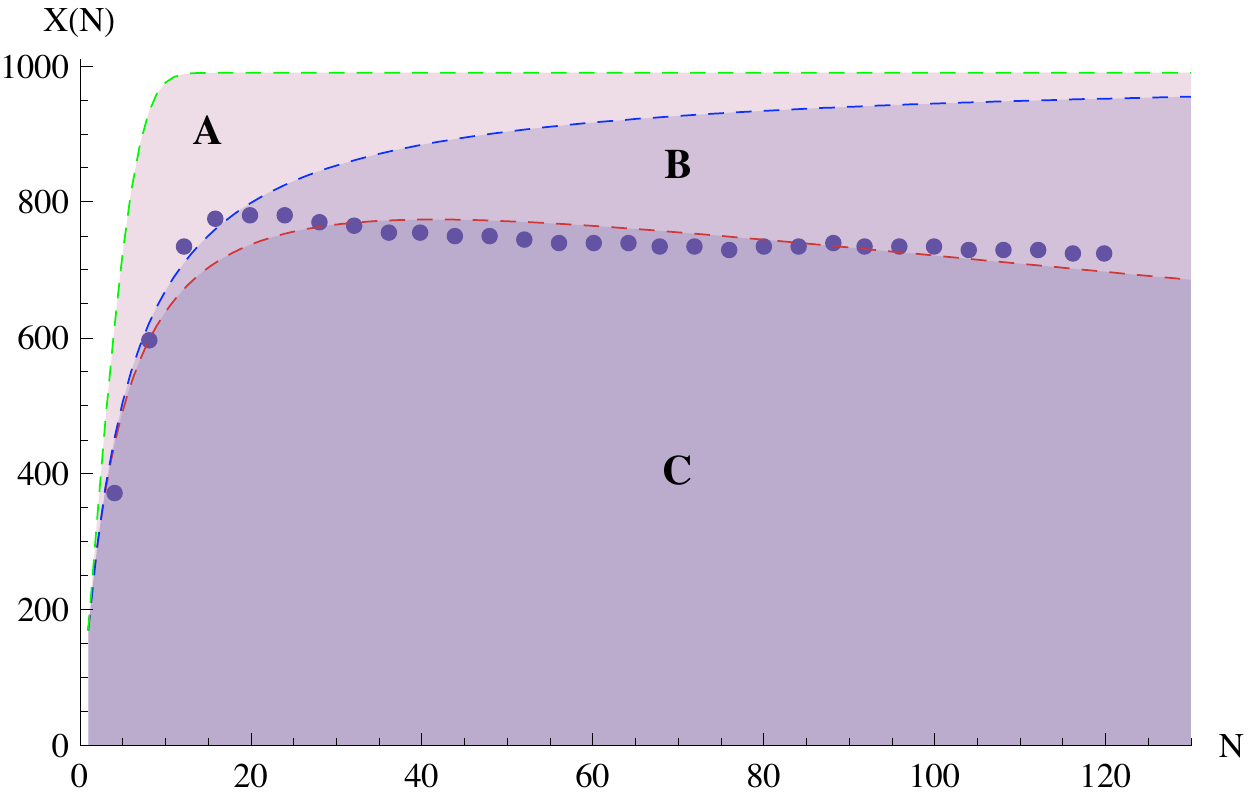} 
\caption{Scalability zones. Unnormalized throughput data $X(N)$ from
Fig.~\ref{fig:USLwebsphere} overlaid on scalability zones denoted
$A,B$ and $C$ consistent with Table~\ref{tab:appclass}}
\label{fig:zones}
\end{figure}

Figure~\ref{fig:figure9} shows the USL model run with the Parallel
service time deterministic (FIXED 0.9) and the Serial service time a
normal distribution (NORMAL(0.1, 0.02), i.e., a mean of $0.1$ and a
standard deviation of $0.02$.  The result is very similar to
Fig.~\ref{fig:simuslx}.
 
\ssection{SCALABILITY ZONES} \label{sec:zones}
The simulation results presented in the previous section, not only verify 
theorem~\ref{thm:usl} for the USL, but they also 
provide a deeper insight into the nature of 
the ``Three C's'' in Sect.~\ref{sec:paramusl}. In fact, we could 
reinterpret them as follows:

\begin{enumerate}\addtolength{\itemsep}{-0.5\baselineskip}
\item \fbox{C}oncurrency-limited scalability   
\mbox{($\alpha,\beta = 0$)} corresponds to \uline{asynchronous} queueing at the 
repairman, which is the same as the mean value solution (\ref{eqn:mrmx}).
\item \fbox{C}ontention-limited scalability  
\mbox{($\alpha > 0, \beta = 0$)} corresponds to synchronous queueing at the 
repairman.
\item \fbox{C}oherency-limited scalability \mbox{($ \alpha,\beta > 0$)}
corresponds to synchronous queueing at a prepping repairman.
\end{enumerate}
The particular meaning ascribed to the word ``repairman'' can be decided
upon using Table~\ref{tab:mrm}. Moreover, as described in
Sect.~\ref{sec:paramusl}, each of the C's is associated with a term in
the denominator of the USL equation and, taken separately, each of them
corresponds to a distinct scalability curve: (1) concurrent linearity,
(2) synchronous contention (Amdahl's law), and (3) synchronous
contention with load-dependent service. These curves are shown as dashed
lines in Fig.~\ref{fig:zones}. The usual convention is to focus on only
one of these possible curves to assess scalability. We propose, instead,
to consider the three regions {\em between} these curves as defining
three scalability zones (See Fig.~\ref{fig:zones}):
\begin{description}\addtolength{\itemsep}{-0.5\baselineskip}
\item[Zone A:] Linear scalability zone associated with asynchronous requests.
\item[Zone B:] Amdahl-limited scalability zone associated with synchronous requests.
\item[Zone C:] Coherency-limited scalability associated with 
synchronous requests and exchange-dependent service.
\end{description}

Just as water is only one of three possible phases (viz., ice, water, 
steam) that exists in a particular temperature range 
\mbox{$(32^{\circ} F < T < 212^{\circ}F)$}, 
so an application can exist in any one of the three
performance phases or zones A, B or C, for a given range of user-loads
($N$). Similarly, just as water undergoes a {\em phase transition} to
steam (i.e., boiling) with increasing temperature ($T \geq 212^{\circ}F$), so
an application can transition between zones as a function of increasing
load.

Figure~\ref{fig:zones} presents a case in point. At low load, the
WebSphere data lies in Zone A which is bounded by concurrent linearity
on the upper side and synchronous contention on the lower side. The data
is closer to the lower bound for $N < 20$. The new interpretation of these data is
that synchronous queueing appears to dominate scalability at low loads.
At $N=20$, just like ice melting, the behavior of WebSphere changes
rather dramatically as it starts to transition across Zone B 
and onto the upper side of Zone C. Above $N \geq 40$, WebSphere  
oscillates along the boundary between Zones C and B.

We know from our extended MRM model that Zone C means load-dependent
service is superimposed on top of synchronous messaging in Zone B. This
could be occurring for many reasons. Some examples are: priority sorting
of the message queue, garbage collection or undesirable memory leaks. We
cannot provide a definite explanation without a more thorough
investigation of the architecture but that is the job of the software 
architect or the software engineer, not the performance analyst. 

However, we can help the software architect or engineer by directing
their attention to the $\mathcal{O}(N^2)$ performance degradation in the
USL model and explaining that Zone C involves synchronous requests with
exchange-dependent service. In this way, the zones interpretation can
quickly narrow the range of potential causes; something that would
otherwise be very difficult to do. As implausible as this may seem to 
the casual reader, it is our experience that hints of this type 
are more than sufficient to trigger ``Eureka!'' moments among those 
most knowledgeable about the architectural details.

Another practical insight, that emerges from our Zones view, is also
worth noting. The zones in Fig.~\ref{fig:zones} suggest one way to
improve throughput performance, viz., attempt to replace the synchronous
messaging, evident in Zones B and C, with asynchronous messaging. This
strategy is analogous to the well-known performance gains that can be
achieved by replacing synchronous (blocking) I/O with asynchronous
(non-blocking) I/O. See
\href{http://en.wikipedia.org/wiki/Asynchronous_I/O}{wiki/Asynchronous\_I/O}.

Clearly, asynchronous messaging should make throughput scale almost linearly 
(Gustafson's dream of Sects.~\ref{sec:paramgus} and~\ref{sec:simgus}) but only for
very low loads. Beyond $N \simeq 10$ users, the throughput 
reaches saturation and (\ref{eqn:mrmr}) tells us that user
response-time will begin to climb up the proverbial ``hockey stick''
handle. However, such linearity may not be desirable from a
performance management perspective. It may be preferable to reach saturation more slowly
and accommodate more aggregate users.  Since this is tantamount to
keeping on the upper side of Zone B, it is only necessary to ensure that
the prepping repairman effect be minimized in the application.
It is not necessary to eliminate synchronous messaging.

\ssection{CONCLUSION}
In this paper, we have used event-based simulation as an exploratory 
tool to accomplish several things. Simulation has confirmed the USL
parametric modeling equation as being physical in the sense that it
corresponds to the synchronous bound on throughput in a particular
queueing model: a prepping machine repairman (Sect.~\ref{sec:simusl}).
This result is the generalization of an earlier theorem concerning a 
queueing interpretation of Amdahl's law based on
rational functions~\citep{arxiv02}.

By virtue of our approach, we have shown that Amdahl and Gustafson
scaling laws are also unified by the {\em same} queueing model, viz.,
the machine-repairman model. Moreover, corollary~\ref{cor:amdmrm} is a
lower bound on throughput; synchronous throughput, and therefore
represents worst-case scalability. With this physical interpretation, it
follows immediately that Amdahl's law can be ``defeated'' more
conveniently than proposed in~\citep{nelson} by simply requiring that
all requests be issued \emph{asynchronously}.

To understand the USL in terms of the machine repairman, the standard
queueing model had to be extended to include: (i) synchronous queueing
(Sect.~\ref{sec:synq}) and (ii) state-dependent service
(Sects.~\ref{sec:prepman}. The precise nature of the synchronous
queueing was only revealed by simulation, because the analytic equations
used in the proof of theorem~\ref{thm:usl} are steady-state equations.
Consequently, they hide the details of how the synchronization occurs, as well as
obscuring how it controls the possible statistical distributions of the
$S$ and $Z$ times in Table~\ref{tab:mrm}.

The simulation models provide a more intuitive understanding of how all
these effects combine in a non-mathematical way. They also reveal how
``real world'' applications might behave (Table~\ref{tab:appclass}) and how
this behavior is reflected in the parametric models used for statistical
regression (Table~\ref{tab:appclass}). It is this concrete physical
interpretation of the USL regression parameters that make it a more
practical tool than the traditional queueing-model approach for
assessing application scalability.

Finally, our investigations have led us to abandon the usual goal of
fitting any particular nonlinear scalability model to data. Rather, we
treat the data as dynamic and thus capable of making transitions between
scalability zones (Sect.~\ref{sec:zones}) as a function of load. Each of
these zones comes with a well-defined interpretation in terms of
queueing effects and this can be vital for system architects and
performance engineers when considering how to get into a better
scalability zone.

\ssection{\large ACKNOWLEDGMENTS}
One of us (NJG) thanks Guerrilla-graduates Denny Chen and 
Paul Puglia for helpful remarks that inspired the proofs of 
theorems~\ref{thm:universal} and~\ref{thm:usl}.

%%%%%%% REFERENCES %%%%%%%%%%
\setlength{\bibsep}{1pt}    % reduce vertical space b/w bibitems
\bibliography{cmg8075}
\bibliographystyle{alpha}

\section*{\sf \large \textbf{TRADEMARKS}}
JavaOne is a service mark of Sun Microsystems, Inc.
Mathematica is a registered trademark of Wolfram Research, Inc. 
SIMUL8 is a registered trademark of SIMUL8 Corporation.
WebSphere is a registered trademark of IBM Corporation.
All other Trademarks, product and company names 
are the property of their respective owners.

\end{document}